\documentstyle[aps,pra,subfigure,epsfig,amsmath,amssymb]{revtex}

\begin{document}
\draft

\title{Particle density and non-local kinetic energy density functional for
two-dimensional harmonically confined Fermi vapors} 

\author{Anna Minguzzi$^{*}$, N.~H. March$^{\dagger,\S}$ and
M.~P. Tosi$^{*}$} 

\address{$^{*}$ Istituto Nazionale per la  Fisica della Materia and
Classe di Scienze, Scuola Normale Superiore, \\ Piazza dei Cavalieri 7,
56 126 Pisa, Italy \\
$^\dagger$  Oxford University, Oxford, England \\
$^\S$ Department of Physics, University of Antwerp (RUCA),
Antwerp, Belgium \\
}

\maketitle                
\begin{abstract}
We evaluate analytically some ground state properties of
two-dimensional harmonically confined Fermi vapors with
isotropy and for an arbitrary number of closed shells. 
We first derive a differential form of the
virial theorem and an expression for the 
kinetic energy density in terms of the fermion particle
density and its low-order derivatives.
These results allow  an explicit  
differential equation to be obtained   for the
particle density. The equation is third-order, linear and homogeneous.
We also obtain a relation between the turning points of kinetic energy
and particle densities, and an expression of the non-local kinetic
energy density 
functional.
\end{abstract}


\section{Introduction}
We have previously studied in detail the ground-state properties of
harmonically confined non-interacting fermions in one dimension
\cite{noi,march_pra}, because of current interest in experiments on
ultracold gases of the $^{40}$K and $^6$Li fermionic isotopes
populating hyperfine states inside
magnetic traps \cite{jila_exp,paris_exp}.  The $s$-wave collisions between
fermions in the same hyperfine state are suppressed by the Pauli
principle, while $p$-wave scattering and dipole-dipole  magnetic
interactions are very weak at very low temperature \cite{fermini}. A
one-component gas of Fermi atoms in a fully spin-polarized state
inside a magnetic trap thus is a close laboratory realization of
the ideal non-interacting Fermi gas in an external harmonic potential,
allowing access to the kinetic energy functional which is invoked in
the density functional theory of inhomogeneous fluids \cite{dft}.  

The interest in such experiments has motivated us to study by
essentially analytical methods the extension of the results of
Ref.~\cite{noi} and \cite{march_pra} to treat independent fermions
with two-dimensional (2D) isotropic harmonic confinement, for an arbitrary
number of closed shells. Throughout we study singly filled levels
because of the experimental situation involving spin-polarized
fermions.

The outline of the paper is then as follows. In Sec.~\ref{sec2} below, we
derive an explicit form of the kinetic energy density for 2D harmonic
confinement for an arbitrary number of closed shells. In Sec.~\ref{sec3} a
differential form of the usual integral virial theorem, but now
specific to 2D harmonic confinement, is invoked and this allows the
kinetic energy density result of Sec.~\ref{sec2} to be rewritten as a
differential equation for the particle density. Sec.~\ref{sec4} gives
the derivation of a relation between the turning points of the
kinetic energy and particle densities. This allows the construction 
of an explicit form of the kinetic energy density functional solely in
terms of the particle density and its low-order  
derivatives. Illustrative numerical results are presented in
Sec.~\ref{sec6}. The final section constitutes a summary, together with some
proposals for further studies.

\section{Derivation of the relation between average kinetic energy density
 and particle density}  
\label{sec2}

Three forms of the kinetic energy density are used in the paper: {\it (i)}
$t(r)=-(\hbar^2/2m) \sum_i \psi_i \nabla^2 \psi_i$, {\it (ii)}
$t_G(r)=(\hbar^2/2m) \sum_i |\nabla \psi_i|^2$ and {\it (iii)} an average of
$t(r)$ and $t_G(r)$ called $\bar t(r)$.  The quantity  $2\bar t(r)$
equals the trace of the  tensorial momentum flux
density and the difference $t_G(r)-t(r)$ is simply $(\hbar^2/4m)\nabla^2
\rho(r)$ where $\rho(r)$ is the particle density. All three kinetic
energy densities defined above lead to the correct total kinetic energy
$T=\int d^2r \, t(r)$ etc.
A brief reminder of the solution of the Schr\"odinger equation for
the problem of present interest is given in Appendix~\ref{appA}.

Our starting point is the result for the 2D fermion particle density
in the presence of the  harmonic confinement $V(r)=m\omega^2
r^2/2$, as 
obtained by Brack and van Zyl \cite{brack}, namely 
\begin{equation}
\rho(r)=\frac{1}{\pi a_{ho}^2}\sum_{\mu=0}^M (M+1-\mu)(-1)^\mu
L_\mu(y) \exp(-y/2)\
\label{rho}
\end{equation} 
after adaptation  to the case of single occupancy of the trap
levels. Here $a_{ho}=(\hbar/m\omega)^{1/2}$, $y=2 (r/a_{ho})^2$,
$(M+1)$ is the number of filled shells and $L_\mu(y)$ are the Laguerre
polynomials. The average kinetic energy density $\bar t(r)$ is given
by \cite{brack}
\begin{equation}
\bar t(r)=\frac{\hbar \omega}{2 \pi a_{ho}^2} \sum_{\mu=0}^M
(M+1-\mu)^2 (-1)^\mu L_\mu (y) \exp(-y/2). 
\label{bart}
\end{equation}

We now define the following summations over the Laguerre polynomials:
\begin{equation}
Q_i^{(M+1)}(y)=\sum_{\mu=0}^M \mu^i (-1)^\mu L_\mu (y): \;\;\;\;\;\;\; i=0,1,2.
\label{qui}
\end{equation}
Then the particle density and the average kinetic energy density are
readily expressed in the form 
\begin{equation}
\rho(r)=\frac{1}{\pi a_{ho}^2}\left[(M+1)
Q_0^{(M+1)}(y)-Q_1^{(M+1)}(y)\right] \exp(-y/2)
\label{rhoqui}
\end{equation}
and
\begin{equation}
\bar t(r)=\frac{\hbar \omega}{2\pi a_{ho}^2}\left[(M+1)^2
Q_0^{(M+1)}(y)-2 (M+1) Q_1^{(M+1)}(y) +Q_2^{(M+1)}(y)\right] \exp(-y/2)\;.
\end{equation}

The next step is to utilize the differential equation for the Laguerre
polynomials, namely 
\begin{equation}
\left[y \frac{d^2}{dy^2}+(1-y) \frac{d}{dy} +\mu\right]L_\mu(y)=0\;.
\end{equation}
This is readily shown to lead to the following relations for the
summations $Q_i^{(M+1)}(y)$:
\begin{equation}
\left[y \frac{d^2}{dy^2}+(1-y) \frac{d}{dy}\right] Q_i^{(M+1)}(y)
+Q_{i+1}^{(M+1)}(y)=0 
\end{equation}
for $i=0$ and 1. These allow us to express $\rho(r)$ and $\bar t(r)$
solely in terms of $Q_0^{(M+1)}(y)$ and of its first two derivatives,
leading after some 
manipulations to the result
\begin{equation}
\bar t(r)=\frac{\hbar \omega}{2}\left[\frac{a_{ho}^2}{8} \nabla^2
\rho(r) +\left(M+\frac{3}{2}-\frac{r^2}{2 a_{ho}^2}\right)\rho(r)\right]\;.
\label{tvsrho}
\end{equation}
Equation~(\ref{tvsrho}) is one   of the central relations for the
present study.

\section{Differential equation for fermion particle density}
\label{sec3}
Having related the kinetic energy density $\bar t(r)$ to the
particle density $\rho(r)$ and the potential energy $m\omega^2 r^2/2$
in Eq.~(\ref{tvsrho}), for an arbitrary number of closed shells, we shall
next invoke a differential form of the usual integral virial theorem.
This we write in the form:
\begin{equation}
\frac{\partial\bar t(r)}{\partial r}=-\rho(r) \frac{\partial
V(r)}{\partial r}\;.
\label{virdiff}
\end{equation}
For the specific case of 2D harmonic confinement with filled shells, 
the differential virial theorem~(\ref{virdiff}) can be proven 
starting from Eqs.~(\ref{rho}) and~(\ref{bart}) for any  given number 
of shells. The demonstration is given in Appendix~\ref{appB}. 

Forming the virial ${\mathbf r }\cdot {\mathbf F}$ of the force
${\mathbf F}$, as $-r \partial V(r)/\partial r$, and integrating over
$2 \pi r dr$ we find
\begin{equation}
\int_0^\infty 2\pi r^2 \frac{\partial \bar t(r)}{\partial r}
dr=\langle {\mathbf r}\cdot{\mathbf F}\rangle = \int d^2 r \rho(r) r
\frac{\partial 
V(r)}{\partial r}\;.
\label{virint}
\end{equation}
Integrating the LHS of Eq.~(\ref{virint}) by parts we obtain $2T$
for the integral value, $T$ being the total kinetic energy, which
confirms that Eq.~(\ref{virdiff}) contains the integral virial theorem.

Having established Eq.~(\ref{virdiff}),  we can combine it with
Eq.~(\ref{tvsrho}) relating 
$\bar t(r)$ to $V(r)$ and $\rho(r)$ to obtain a
a third-order, linear, homogeneous differential equation for the
particle density $\rho(r)$ for $(M+1)$ closed
shells: 
\begin{equation}
\frac{\hbar^2}{8m} \frac {\partial}{\partial r} \left[\nabla^2
\rho(r)\right]+\left[\left(M+\frac{3}{2}\right)\hbar
\omega-\frac{m\omega^2 r^2}{2}\right]\frac{\partial \rho(r)}{\partial
r}+ m\omega^2 r \rho(r)=0\;. 
\label{eqdiff}
\end{equation}
As one immediate check of Eq.~(\ref{eqdiff}), it is readily
verified that 
\begin{equation}
\rho(r)= {\cal N} \exp(-r^2/a_{ho}^2)\;,
\end{equation}
where ${\cal N}$ is a known (irrelevant for present purposes)
normalization factor, satisfies Eq.~(\ref{eqdiff}) for $M=0$.

Equation~(\ref{eqdiff}) achieves a long-term aim of density functional
theory for the present case of 2D  harmonic confinement of
independent fermions: namely to allow the direct calculation of the
particle density from the potential energy ($m\omega^2 r^2/2$ in this
case), without recourse to individual wave functions. It constitutes
the two-dimensional generalization of the result of Lawes and March
\cite{lawes} for one-dimensional (1D)  harmonic confinement of fermions.

\section{Kinetic energy functional}

\label{sec4}

The first object of the present section is to obtain a relation between
 $t'(r)=\partial t(r)/\partial r$ and $\rho'(r)=\partial \rho(r)/\partial
r$ for this isotropic
2D harmonically confined fermion assembly. 
The result for 1D harmonic confinement is 
(see {\it e.g.} \cite{march_pra})
\begin{equation}
\frac{t'(x)}{\rho'(x)}= \left(N \hbar \omega-\frac{m \omega^2 x^2}{2}\right)
\label{turning1d}
\end{equation}
We first use the definition of $\bar t(r)$ and the
relation $t_G(r)=t(r)+(\hbar^2/4m)\nabla^2 \rho(r)$ to obtain
\begin{equation}
t (r)=\bar t(r)-\frac{\hbar^2}{8m} \nabla^2 \rho(r)\;.
\label{tbart2}
\end{equation}
After differentiation of Eq.~(\ref{tbart2}) with respect to $r$, the
use of the virial theorem~(\ref{virdiff}) and of the differential
equation~(\ref{eqdiff}) for $\rho(r)$  allows us to obtain
\begin{equation}
\frac{t'(r)}{\rho'(r)}= 
\left[\left(M+\frac{3}{2}\right)\hbar \omega -\frac{m \omega^2
r^2}{2}\right] \;,
\label{turning}
\end{equation}
which is the desired generalization of Eq.~(\ref{turning1d}) to
2D  harmonic confinement. 
Evidently Eq.~(\ref{turning})
 relates directly the turning points of
kinetic energy and particle densities for any chosen number of
closed shells. 

Equation~(\ref{turning}) allows us to eliminate the presence of the
external potential and to obtain the non-local kinetic
energy density functional solely in terms of the particle density and of its
low-order derivatives. To this aim, we  first write an
equation for the quantity $Q(r)=t'(r)/\rho'(r)$ by differentiation of
Eq.~(\ref{turning}) and  by using  Eqs.~(\ref{virdiff}) and
(\ref{tbart2}):
\begin{equation}
\frac{\partial}{\partial
r}\left[\frac{Q(r)}{\rho(r)}\right]=\frac{\hbar^2}{8m}\frac{1}{\rho^2(r)}
\frac{\partial}{\partial r} \nabla^2 \rho(r)\;.
\end{equation}
As a second step, we obtain the function $Q(r)$ by integration:
\begin{equation}
Q(r)=\frac{(M+3/2)\hbar \omega}{\rho(0)} \rho(r)+\frac{\hbar^2}{8m}
\rho(r)\int_0^r ds \frac{1}{\rho^2(s)}\frac{\partial}{\partial s}\nabla^2
\rho(s)
\;,
\end{equation}
the integration constant being fixed by the condition
$Q(0)=(M+3/2)\hbar \omega$. A further integration and elimination of
the high-order derivatives through integration by parts leads to the
final expression
\begin{equation}
t(r)=\dfrac{t_W(r)}{2}+
\left[C+\frac{\hbar^2}{16 m} \int_0^r ds
\frac{[\rho'(s)]^2}{\rho^3(s)}\left(\frac{2}{s}
+\frac{3\rho'(s)}{\rho(s)}\right)\right] \rho^2(r) \;,
\label{func}
\end{equation}
where 
$t_W(r)=(\hbar^2/8 m)[\rho'(r)]^2/\rho(r)$ is the von Weizs\"acker
``surface''  
contribution  to the  kinetic energy
density \cite{weis} and the constant $C$ is given by 
$C=(M+3/2)\hbar\omega/(2\rho(0))-(\hbar^2/16m)(\nabla^2
\rho(r))_{r=0}/\rho^2(0)$. 
This provides a generalization of the result obtained by March {\it et
al.} \cite{vandoren} to the 2D harmonic confinement. 
By integration of Eq.~(\ref{func}) over the system one obtains an explicit
expression for the kinetic energy functional.

\section{Illustrative numerical results}
\label{sec6}

Equations~(\ref{rho}) and~(\ref{bart}) can be employed to numerically
evaluate all the ground state properties defined in Sec.~\ref{sec2}.

\begin{figure}
\centerline{\psfig{file=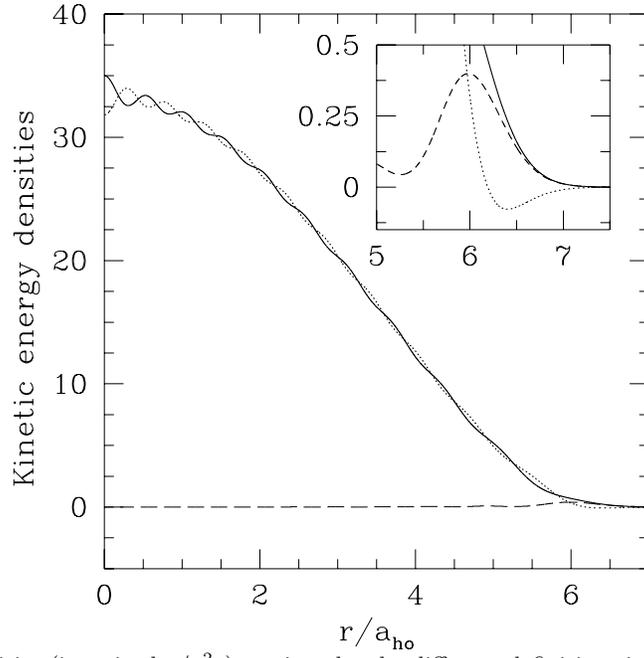,width=0.5\linewidth}}
\caption{Kinetic energy densities (in units $\hbar \omega/a_{ho}^2$)
as given by the different definitions in Secs.~\ref{sec2}
 and~\ref{sec4} for a 2D Fermi 
gas in isotropic harmonic confinement  with 20 filled shells, as functions of
the radial coordinate $r/a_{ho}$. Solid line: $t_G(r)$, dotted line:
 $t(r)$, dashed line: $t_W(r)$. The inset shows an enlargement of the
 turning point region, in the same units.}
\label{fig1}
\end{figure}

In Fig.~\ref{fig1} we plot the exact kinetic energy densities $t(r)$
and  $t_G(r)$
for 20 filled shells ($M$=19). There
is seen to be a tiny negative region of $t(r)$, whereas $t_G(r)$  is by
definition everywhere positive. 
In the same figure  we  also show the von Weizs\"acker    
kinetic energy
density $t_W(r)$.
It is seen that $t_W(r)$ is practically zero until the
classically forbidden region 
is reached, while well outside  the
classically allowed region it eventually merges
with $t_G(r)$: 
{\it i.e.} $t_G(r)\rightarrow t_W(r)$ for $r$ greater than the Fermi
radius. This property is shown in the inset of Fig.~\ref{fig1}.
 
\begin{figure}
\centerline{\psfig{file=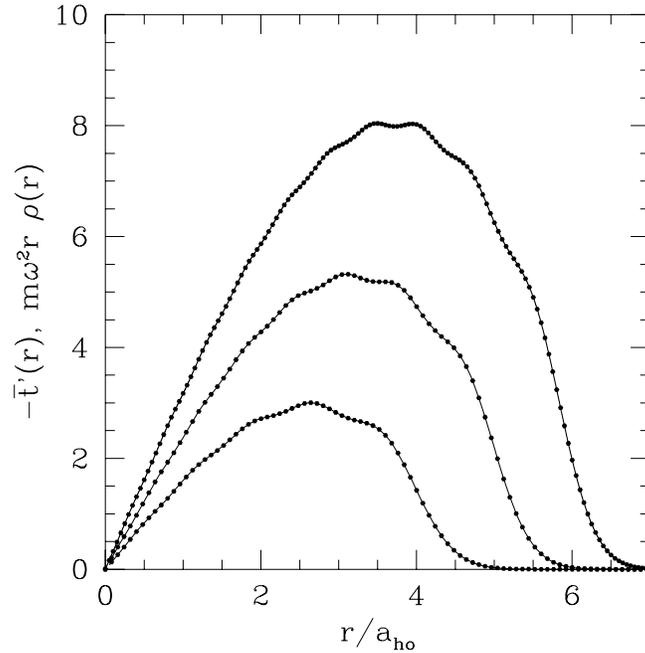,width=0.5\linewidth}}
\caption{Illustration of the differential virial theorem: negative of
 $\partial \bar t(r)/\partial r$ (solid line) and $m\omega^2 r
 \rho(r)$ (dots) in units $\hbar \omega/a_{ho}^3$
 as functions of the radial coordinate $r/a_{ho}$, for a 2D Fermi
gas in isotropic harmonic confinement  with 20, 15 and 10  filled
 shells (from top to bottom).  }
\label{fig2}
\end{figure}

In Fig.~\ref{fig2} the differential virial theorem~(\ref{virdiff})
is illustrated in the case of harmonic confinement 
by comparing  the negative of $\partial \bar t/\partial r$  with the
quantity $m  \omega^2 r \rho(r)$ for various numbers of filled shells. 

\section{Summary and proposed future directions}
We have derived a set of analytic relations for the ground state
properties of a 2D Fermi gas under harmonic confinement,
with the proviso that the Fermi particles always fill $(M+1)$ closed
shells. Our work was aimed mainly at transforming the description of the
system from one given in terms of wave functions to  one  which takes as
the main variable the particle density. 
This procedure, typical of
Density Functional Theory (DFT), is here possible in an  explicit way due to
the simplicity of the system, but may serve as a
guideline to the use of  DFT in more complex systems.

Five basic equations have been obtained, namely Eqs.~(\ref{tvsrho}),
(\ref{virdiff}), (\ref{eqdiff}), (\ref{turning}) and~(\ref{func}).
Equation~(\ref{tvsrho}) is a differential relation which gives the
average kinetic energy density in terms of the particle density, and
Eq.~(\ref{virdiff}) is a differential version of the usual virial
theorem. These relations  allow the derivation of
Eq.~(\ref{eqdiff}),  a third-order differential equation for the
particle density profile as determined by the external confining
potential and by the number of filled shells, and Eq.~(\ref{turning}), a
relation between the turning points of the kinetic energy density and
the particle density. The latter is finally used to 
obtain the kinetic energy density functional (\ref{func}) in
terms of the 
particle density,  which is a main task in DFT.

What seems to us therefore to remain outstanding in 2D is to solve the
differential equation~(\ref{eqdiff}), hopefully analytically and
without summation over shells. The motivation for expecting this is
the ``folklore'' that the properties of a fermion fluid are determined
by its highest occupied level \cite{lawes}.
Of course, the challenge of the
three-dimensional oscillator problem remains as a high priority for
further study.

\acknowledgements
This work was partially supported by MURST through
PRIN2000. N.H.M. wishes to acknowledge generous support from the
Scuola Normale Superiore, during the stay in Pisa in which his
contribution to the present study was brought to fruition. 
N.H.M. acknowledges valuable discussions in the general area embraced
by this article with Dr. L.~M. Nieto and Professors l.~C. Balb\'as,
A. Holas and F.~M. Peeters. Finally, A.M. acknowledges discussions with
Dr. P. Tilli. 

\appendix
\section{Two-dimensional Fermi gas in harmonic confinement: wave
functions and Dirac density matrix}
\label{appA}

The Schr\"odinger equation in 2D reads
\begin{equation}
-\frac{\hbar^2}{2m} \nabla^2 \psi+\frac{1}{2} m \omega^2 r^2
 \psi=\varepsilon \psi\;.
\label{schroed}
\end{equation}
Using $a_{ho}\equiv (\hbar/m\omega)^{1/2}$ and $\hbar \omega$ as units
of length and energy, the eigenfunctions of Eq.~(\ref{schroed}) are
expressed through the Laguerre polynomials $L_n^m(r^2)$ as
\begin{equation}
\psi_{n,m}(r,\phi)=\left(\frac{n!}{\pi (n+m)!}\right)^{1/2} r^m
L_n^m( r^2) \exp(-r^2) \exp(i m\phi)\;,
\label{psinm}
\end{equation}
in correspondence to the energy eigenvalues 
\begin{equation}
\varepsilon_{n,m}=2 n+m+1\;.
\end{equation}
The various shells are filled by taking all integer values of $n$ and
$m$ which satisfy $2n+m=N$, $N$ being the shell index.

The ``off-diagonal density'', that is the Dirac density matrix can be
immediately written from Eq.~(\ref{psinm}) as
\begin{equation}
n({\mathbf r,r'}) =\frac{1}{\pi}
\exp\left[-(r^2+{r'}^2)/2\right]\sum_{N=0}^{M} \sum_{h=0}^N
\frac{(N-h)!}{h!} (r r')^{2h-N} L_{N-h}^{2h-N}(r^2)L_{N-h}^{2h-N}({r'}^2)\exp[i(2h-N)(\phi-\phi')]\;.
\label{dirac}
\end{equation}
From this matrix, the density profile $\rho(r)$ for $(M+1)$ filled shells
is obtained by taking ${\mathbf r'=r}$, the result being
the same as the one obtained by Brack and van Zyl
\cite{brack}.  Similarly, the kinetic energy density can be directly
constructed in either the form of $t(r)$ or $t_G(r)$ by appropriate
differentiation on the Dirac matrix~(\ref{dirac}), and afterwards
allowing ${\mathbf r'}$ to tend to ${\mathbf r}$. 

\section{Proof of the differential virial theorem for a 2D
harmonically confined Fermi gas}
\label{appB}

We demonstrate here Eq.~(\ref{virdiff}) for the specific case 
$V(r)=m\omega^2 r^2/2$, which is the one relevant for the present
work. The units $a_{ho}$ and $\hbar\omega$ for length and energy are
used throughout. 

As a first step, we derive from Eqs.~(\ref{rho}) and Eq.~(\ref{bart})
recursion relations for the 
particle density $\rho_{M+1}(r)$ and the 
average kinetic energy  density $\bar t_{M+1}(r)$  of a system with $(M+1)$
filled shells in terms of those of a system with $M$ filled
shells. These relations are
\begin{equation}
\rho_{M+1}(r)= \rho_M(r)+Q_0^{(M+1)}(y) \exp(-y/2)/\pi
\end{equation}
and
\begin{equation}
\bar t_{M+1}(r)= \bar t_M(r)+\rho_{M}(r)+Q_0^{(M+1)}(y) \exp(-y/2)/(2\pi)\;.
\end{equation}
These involve the function $Q_0^{(M+1)}(y)$ defined in
Eq.~(\ref{qui}).

We then proceed by induction. The differential virial theorem is
readily proven for the case $M=1$ of a singly occupied shell by
direct substitution of the expressions $\rho_1(r)=\exp(-r^2)/\pi$ and
$\bar t_1(r)=\exp(-r^2)/2\pi$ into Eq.~(\ref{virdiff}). Assuming that
the theorem holds for $M$ filled shells, we are left to prove that 
\begin{equation}
\frac{\partial}{\partial r}\left[\rho_{M+1}(r)-\frac{e^{-y/2}}{2\pi}Q_0^{(M+1)}(y)\right]=-r\frac{e^{-y/2}}{\pi} Q_0^{(M+1)}(y)\;.
\label{ecco}
\end{equation}
Physically this means that the theorem holds separately for
each filled shell. 

By the use of Eq.~(\ref{rhoqui}) for the density
profile, Eq.~(\ref{ecco}) can be rewritten solely in terms of the
functions $Q_i^{(M+1)}(y)$ with $i=0,1$:
\begin{equation}
\left(M+\frac{1}{2}\right)\frac{\partial Q_0^{(M+1)}(y)}{\partial y} -
2M Q_0^{(M+1)}(y)=\frac{\partial Q_1^{(M+1)}(y)}{\partial y} -
2 Q_1^{(M+1)}(y)\;.
\label{lariuso}
\end{equation}
We can now employ the definition of the Laguerre polynomials
\begin{equation}
L_{\mu}(y)=\sum_{k=0}^\mu \left(\begin{array}{c} \mu \\ \mu-k \end{array}\right) \frac{(-y)^k}{k!}
\end{equation}
and of the functions $Q_i^{(M+1)}(y)$ to transform Eq.~(\ref{lariuso})
into a double sum:
\begin{equation}
\sum_{\mu=1}^M (-1)^\mu\sum_{k=0}^{\mu-1} (2M+1-2\mu)
\left(\begin{array}{c} \mu \\ \mu-k-1 \end{array}\right)
\frac{(-y)^k}{k!}+\sum_{\mu=0}^M (-1)^\mu\sum_{k=0}^{\mu} (M-\mu) 
\left(\begin{array}{c} \mu \\ \mu-k \end{array}\right)
\frac{(-y)^k}{k!}=0\;.
\end{equation}
Exchanging the order of the sums allows us to collect all terms of the
same order in $k$: each coefficient must then vanish independently because
of the principle of identity between polynomials. Thus, we have
finally to prove the simplified expression
\begin{equation}
\sum_{\mu=k}^{M-1}(-1)^\mu \left[(2M-1-2\mu)(\mu+1)-(M-\mu)(k+1)\right]
 \frac{\mu!}{(\mu-k)!}=0 \;. 
\label{larichiamo}
\end{equation}
The above finite sum must be shown to yield zero for any integer value
$M$ and $k\le (M-1)$. This statement is proven by applying twice again
the reasoning by induction on the index $M$: the validity of
Eq.~(\ref{larichiamo}) for $M=1$ is easily checked, and assuming that
it holds for a fixed $M$ leads to prove that 
\begin{equation}
\sum_{\mu=k}^{M-1}(-1)^\mu (2 \mu-k+1)
\frac{\mu!}{(\mu-k)!}=(-1)^{M+1}\frac{M!}{(M-k-1)!} \;.
\label{anchelei}
\end{equation}
Applying again induction to the LHS of Eq.~(\ref{anchelei}) leads to
the proof of its validity, which implies the validity of the
differential virial theorem.

\end{document}